\documentclass[floatfix]{revtex4}
\usepackage{epsf,graphicx}
\usepackage{amsmath,amssymb,amsfonts}
\begin{document}
\title{Reply to Jarzynski's comment cond-mat/0509344 \cite{jarzynski05}}
\author{D.H.E. Gross}
\affiliation{Hahn-Meitner Institute and Freie Universit{\"a}t Berlin,\\
Fachbereich Physik.\\ Glienickerstr. 100\\ 14109 Berlin, Germany}
\email{gross@hmi.de} \homepage{http://www.hmi.de/people/gross/ }
\maketitle              
In ``Flaw of Jarzynski's equality when applied to systems with several
degrees of freedom'' \cite{gross217} I gave a simple microcanonical example
against Jarzynski's equality\cite{jarzynski97}
\begin{equation}
\left<\exp(-\beta W)\right>=  \exp(-\beta \Delta F):\label{jarzynski}
\end{equation}
When applied to an ideal gas of $N$ particles with mass $m$
streaming under thermal isolated conditions (microcanonical) from
a small volume $V_0$ into a larger volume $V_1>V_0$ the relation
(\ref{jarzynski}) is violated. If the piston closing the container
$V_0$ is removed fast, no work is done ($W=0$) but the free energy
decreases by $\Delta F=-NkT\ln(V_1/V_0)$.

Jarzynski correctly replies in his comment that
eq.(\ref{jarzynski}) does not apply to a microcanonical system. It
is only valid for a system equilibrized in a {\em canonical} heat
bath. Even if the piston is pulled out in $x-$direction much
faster than the thermal velocity of the gas particles (mass $m$)
\begin{equation}
{v_x}_{therm} \sim \sqrt{kT/m},
\end{equation} there
are still particles in the far tail of the Maxwell velocity
distribution with high enough velocities to hit the piston
c.f.\cite{lua05,sung05}. Then the gas performs work on the piston
and relation (\ref{jarzynski}) may be fulfilled. This occurs, of
course, only in the canonical ensemble in contrast to the
microcanonical one. In the latter the velocity of any particle is
smaller $v\le \sqrt{2E_{tot}/m}= \sqrt{N 3 kT/m}$ when all other
particles are at rest.

Even though this is formally correct it does not happen in
reality: In a gas of carbon atoms at $T=300^°$ Kelvin the thermal
velocity is ${v_x}_{therm} \sim 500$m/sec. Pulling the piston with
a speed $~10\times {v_x}_{therm}$ allows only atoms with an energy
$\ge 100$kT to collide with it. These events, however, have a
probability of $\mbox{prob}\sim e^{-100}$. The rate of particles
hitting the piston is \cite{reif65}:
\begin{equation}
\frac{1}{\Delta t}=\frac{P\times
\mbox{prob}\times\mbox{area}}{\sqrt{2\pi m kT}}
\end{equation}
If the pressure  of the gas is $P=1$atm., and the piston closing the vessel
$V_0$ has an area of $1$cm$^2$, such a collision will occur only every
$10^{12}$ years! Evidently, in any experiment which Jarzynski has in mind
the gas behaves approximately more {\em microcanonically} with a truncated
velocity distribution than with a canonical one. Then the flaw of
Jarzynski's identity (\ref{jarzynski}) discussed in \cite{gross217} can be
seen.

However, this may be a technical argument against relation
(\ref{jarzynski}). There is also a principal one: As well known,
the canonical Boltzmann-Gibbs statistics is an approximation to
the fundamental microcanonical statistics. Only in the
thermodynamic limit of infinitely many particles it {\em may} be
equivalent to the microcanonical (for the most serious exceptions
see \cite{gross216}). Now, Jarzynski's equality (\ref{jarzynski})
is frequently applied to nano-systems like single DNS
molecules\cite{bustamante05}. Following Hartmann et
al.\cite{hartmann04} I discussed in \cite{gross210} the failure of
assuming a canonical ensemble for a nano-system. Even if such a
small system is {\em equilibrized in a canonical} heat bath, its
energy distribution is {\em not} canonical.

My main argument in  \cite{gross217}, however, was against a
statement in \cite{jarzynski97}, which Jarzynski considers as his
central result:
\begin{equation}
\frac{P_+(z_B,+\Delta S_b|z_A)}{P_-(z_A^*,-\Delta
S_b|z_B^*)}=e^{\Delta S_b}\hspace{2.5cm} \label{jarzcentral}
\end{equation} where $P_+(z_B,+\Delta S_b|z_A)$ is the probability of the
system $\psi$ to go in its phase space $\Psi$ from a point $z_A$
to $z_B$ under the energy transfer $\Delta Q=T\Delta S_b$ to the
bath. $P_-(z_A^*,-\Delta S_b|z_B^*)$ is the probability for the
time reversed process of $\psi$ to go from the point
$z_B^*=\{q_B,-p_B\}$ in phase space $\Psi$ back to the point
$z_A^*=\{q_A,-p_A\}$ conjugate to the initial point $z_A$, and
with the opposite energy transfer $-\Delta Q$.

If the system $\psi$ is complex and has several degrees of freedom
(e.g. a DNA molecule, or our ideal gas above), then the phase
space $\Psi$ is high dimensional and the path from $z_A$ to $z_B$
is not so simple to revert. All velocities at $z_B$ must
simultaneously be reverted exactly. This is analog to Loschmid's
conjecture against Boltzmann. Normally, this is not possible in
practice. Here Clausius comes into play who clearly stated already
in \cite{clausius1854}: The entropy created by the whole process
is
\begin{equation}
\Delta S>\Delta S_b=\Delta Q/T,
\end{equation} when the control of all degrees of freedom
of the system $\psi$ is not perfect. I.e. {\em  more} entropy $\Delta S$ is
generated than the entropy $\Delta S_b$ created in the bath only.
Characteristically, Clausius calls this {\em internal} surplus of entropy
generation the {\em ``uncompensated metamorphosis''} in
\cite{clausius1854}, which must be $\ge0$ in any process. {\em This and
only this is what the second law is about} \cite{prigogine71}. With
$-\beta\Delta F= \Delta S$ in formula (\ref{jarzynski}) the equal sign
becomes a $\le$ - sign.

Concluding: Jarzynski's formula \ref{jarzcentral} may offer an interesting
way to test the fundamentals of statistical mechanics experimentally,
provided some caveats are taken.


\begin{thebibliography}{1}

\bibitem{jarzynski05}
C.~Jarzynski.
\newblock Reply to comments by D.H.E.Gross.
\newblock http://xxx.lanl.gov/abs/cond--mat/0509344, 2005.

\bibitem{gross217}
D.H.E. Gross.
\newblock Flaw of Jarzynski's equality when applied to systems with several
  degrees of freedom.
\newblock  cond--mat/0508721, (2005).

\bibitem{jarzynski97}
C.~Jarzynski.
\newblock {\em Phys. Rev. Lett.}, 78:2690, 1997.

\bibitem{lua05}
Rhonald~C. Lua and Alexander~Y. Grosberg.
\newblock Practical applicability of the Jarzynski relation in statistical
  mechanics: A pedagogical example.
\newblock {\em J. Phys. Chem. B}, 109:6805 --6811, 2005.

\bibitem{sung05}
Jaeyoung Sung.
\newblock Validity condition of the Jarzynski relation for a classical
  mechanical system.
\newblock cond--mat/0506214, 2005.

\bibitem{reif65}
F.~Reif.
\newblock {\em Fundamentals of statistical and thermal physics}.
\newblock McGraw-Hill, New York, 1965.

\bibitem{gross216}
D.H.E. Gross.
\newblock On the foundation of thermodynamics by microcanonical
  thermostatistics.\\ The microscopic origin of condensation and phase
  separations.
\newblock cond--mat/0509202, (2005).

\bibitem{bustamante05}
C.~Bustamante, J.~Liphardt, and F.~Ritort.
\newblock The nonequilibrium thermodynamics of small systems.
\newblock {\em Physics Today}, 58:43, 2005.

\bibitem{hartmann04}
M.~Hartmann, G.~Mahler, and O.~Hess.
\newblock Fundamentals of nano-thermodynamics.
\newblock cond--mat/0408133, 2004.

\bibitem{gross210}
D.H.E. Gross.
\newblock Exploring temperature at the nano-scale.
\newblock {\em Physics World}, October:23, (2004).

\bibitem{clausius1854}
R.~Clausius.
\newblock \"Uber eine ver\"anderte Form des zweiten Hauptsatzes der
  mechanischen W\"armetheorie.
\newblock {\em Annalen der Physik und Chemie}, 93:481--506, 1854.

\bibitem{prigogine71}
P. Glansdorff and I. Prigogine.
\newblock {\em Thermodynamic Theory of Structure, Stability and Fluctuations}.
\newblock{John Wiley\& Sons, London.} 1971

\end{thebibliography}

\end{document}